# Revisiting the Einstein-de Haas experiment: The Ampère Museum's Hidden Treasure

Alfonso San Miguel[1,2] and Bernard Pallandre[2]

(1) Institut Lumière Matière, UMR5306 Université Lyon 1-CNRS 69622 Villeurbanne cedex, France
(2) Musée Ampère, 300 route d'Ampère, 69250 Poleymieux au-Mont-d'Or, France

Unearthed in the Ampère Museum near Lyon, France, a genuine version of the Einstein-de Haas experiment apparatus offers a rare glimpse into Einstein's experimental interests. This remarkable find not only connects us to a crucial epoch in history of science but also highlights Einstein's rare tangible legacy in the realm of experimental physics.

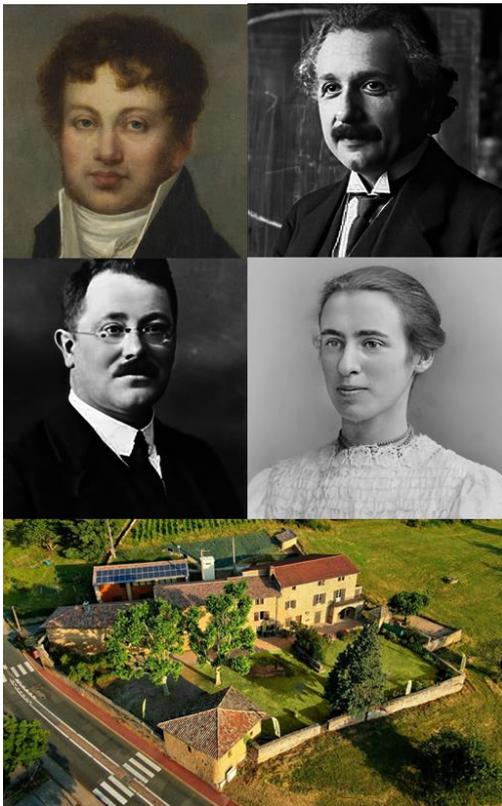

**Fig 1. The main characters of this discovery.** Four physicist (André-Marie Ampère, Albert Einstein Wander Johannes de Haas, and Geertruida Luberta de Haas-Lorentz, wife of W.J. de Haas and daughter of Hendrik Anton Lorentz) and a place: the Ampère Museum (Poleymieux-au-Mont-d'Or, France), which is labeled EPS Historic site.

### Einstein love for experimental physics

Albert Einstein, often hailed as the quintessential theoretical physicist, might astound many with his lesser-known enthusiasm for experimentation. At the age of 35, he revealed in a letter to his friend Michele Besso, "Experimenting is becoming a passion for me" – a statement that peels back the layers of Einstein's scientific persona. This excitement was not just theoretical musing; it was anchored in a series of groundbreaking experiments conducted with Dutch physicist Wander de Haas. Their goal? To demonstrate the existence of Ampère's molecular currents, a venture that represented Einstein's unique involvement in published experimental research, spanning three significant papers on the subject.

In the storage of the Ampère Museum near Lyon, France, we have uncovered a treasure: the only known original and complete Einstein-de Haas experiment apparatus. This remarkable discovery not only links us to a pivotal moment in history of science but also sheds new light on one of Einstein's most significant contributions to experimental physics.

### The Ampère hypothesis

To understand what compelled Einstein to undertake experimental work, we need to look back to the beginning of the 19th century. Ampère, inspired by Oersted's observation of a compass needle's deviation due to a galvanic current, proposed that all magnetic and electric forces were due to current interactions. He theorized that magnetic forces in magnets were due to molecular scale current loops, intuitively hinting at the existence of electrons.

Ampère as well as James Clark Maxwell and many others failed to demonstrate experimentally the existence of Ampère molecular currents. W.J. de Haas and his wife, G. de Haas-Lorentz showed in 1917 that Maxwell's experiments lacked the necessary resolution to evidence Ampère's molecular currents[1].

An analogue problem was found by Owen Willans Richardson[2] who in 1908 devised a similar set-up to the one imagined by Einstein and described below.

**The Einstein-de Haas experiment**

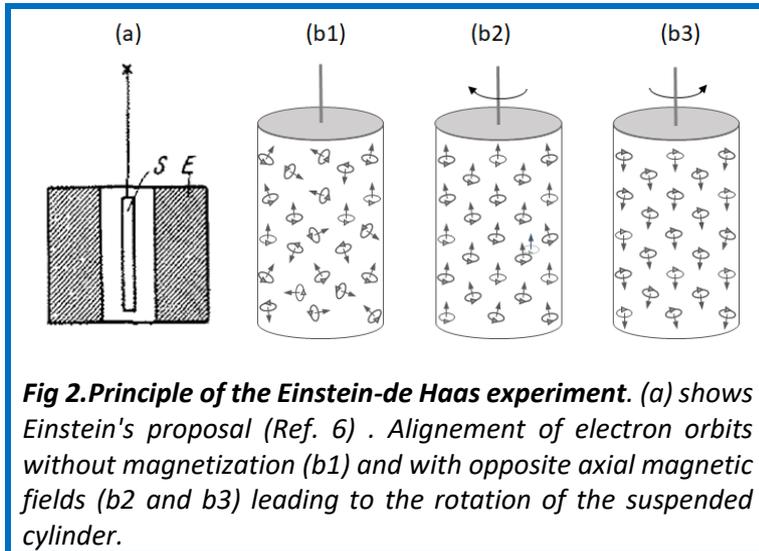

Fig 2.Principle of the Einstein-de Haas experiment. (a) shows Einstein's proposal (Ref. 6) . Alignement of electron orbits without magnetization (b1) and with opposite axial magnetic fields (b2 and b3) leading to the rotation of the suspended cylinder.

In the early 20th century, Ampère's molecular currents were yet to be discovered. Nevertheless, the unveiling of the electron (1897) and the emergence of Bohr atomic model (1913) had strongly modified the physicist view of matter at the "molecular scale". It is in such a context that Einstein and de Haas in 1914–15 undertook an experiment that would test the hypothesis which had resisted demonstration for almost a century.

In the experiment an iron ferromagnetic rod is suspended by a thin wire in the center of a coil (Fig 2.a). The magnetic field generated by a current flowing through the coil which we may consider as initially not magnetized (Fig2.b1) aligns the "Ampère's molecular currents" and their associated magnetic moment magnetizes the bar (Fig2.b2). To comply with the conservation of angular momentum, a torque develops in the bar, causing it to rotate. Inverting the current in the solenoid changes the direction of alignment and consequently the sign of the torque (Fig2.b3). By applying an alternating current at a frequency equal to the natural frequency of mechanical oscillation of the suspended bar an amplified rotation of the iron bar is obtained. This resonant oscillation is a key point which had not been considered by Richardson[2] in his attempts.  An optical mirror fixed on the iron bar acted as a lever arm to amplify the movement of a reflected light source. This method made it possible to observe at the macroscale the effect of Ampère's microscopic hypothesis, enabling the measurement of the gyromagnetic ratio – the ratio of angular momentum to magnetic moment. Under Ampère's hypothesis and the contemporary understanding of physics at the time, both Richardson and Einstein estimated this ratio. They found *2 g e/m*, where *e* and *m* represent the electron's charge and mass, respectively with g=1. The experimental value, reported by Einstein and de Haas was also *g=1*, which was half of what was observed in other experiments. These varying results, with electron *g*-values ranging from *1.0* to *2.2*, highlights the debate and uncertainty within the scientific community at the time, and also emphasizes the delicate and challenging nature of these experiments. The eventual acceptance of the quantum relativistic value, *g=2*, accounting for the electron spin's contribution, was delayed, possibly due to the reluctance to challenge Einstein's findings, given his eminent status in the physics world.

**Einstein involvement and motivations**

Einstein's involvement in conceiving the experiment to demonstrate Ampère's molecular currents is evident from his actions and correspondences. As early as between 1905 and 1909, Einstein may have started experimental attempts in this direction[3]. His inspiration, as he revealed in a letter to Sommerfeld, came while evaluating a gyrocompass patent. The lead of Einstein in this research project is evidenced when he writes to Lorentz on 17 June 1916: "You should not keep thanking me just because I conducted the Ampère investigation with him [*with W.J de Haas*]. For I chose him out of selfishness, because I liked him the best, and my impression turned out to be correct."

In his papers, Albert Einstein cites two main motivations in his experimental work. First, he emphasizes that Ampère's molecular currents concept unifies para- and ferromagnetism, aligning with his goal to integrate diverse physical phenomena. Second, he notes the significance of Curie and Langevin's work on magnetism, particularly how constant molecular magnetic moments at varying temperatures imply electrons' orbital motion at absolute zero, suggesting the existence of zero-point energy, a concept Einstein considered since at least 1913.[4]

**The discovered artifact at Ampère Museum**

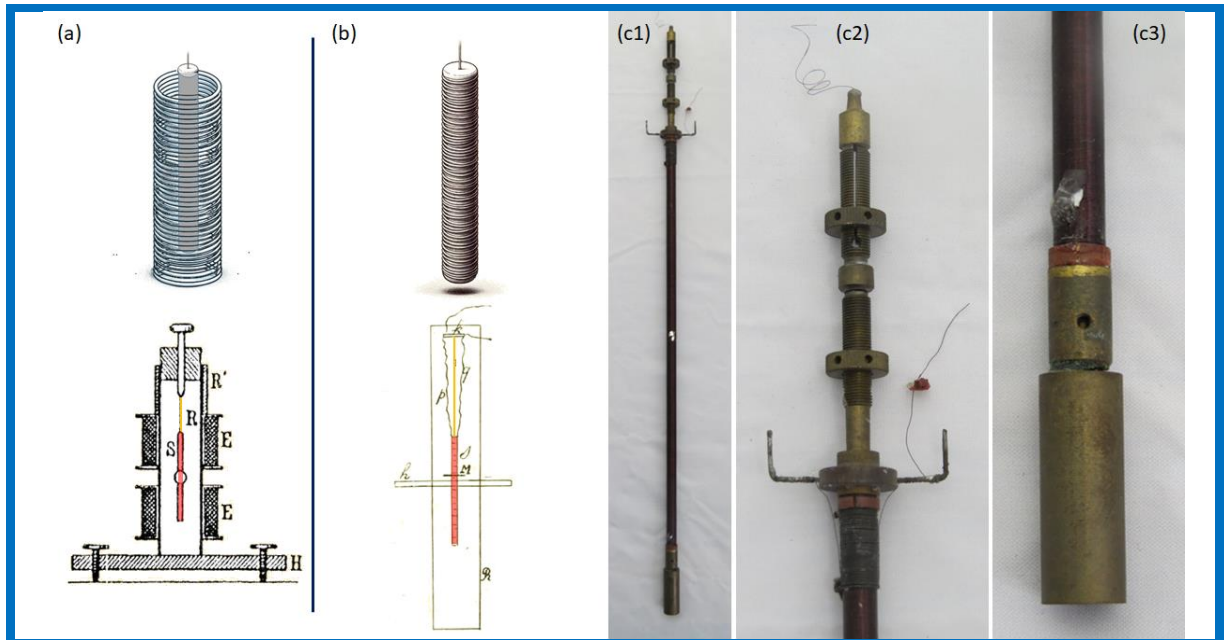

**Fig 3. Principle sketch (top) and original scheme (bottom) of t*he two versions of the Einstein-de Haas experiment:* (a) The Einstein version (Ref. 5); (b) The de Haas version (Ref. 8). (c) is the realization of (b) donated by G. de Haas-Lorentz to the Ampère Museum (c1: full view; c2: upper part; c3: bottom part). The set-up schemes show the iron bar colored in red and the suspension wire in yellow.**

Einstein presented the principles and results of the experiment in three different papers between February and May 1915. He signed alone two of the manuscripts[5,6] and the third and more detailed one together with Wander J. de Haas[7]. The three papers were entitled "Experimental Proof of Ampère's Molecular Currents" or close variants. The principle and set-up of their experiment is shown in Fig3.a. In September 1915, W.J. de Haas presented an improved version of the experimental set-up[8] which is shown in Fig 3.b.

The set-up of de Haas tried to solve various experimental problems that Einstein and de Haas detected in their first experiments. This include the effect of the earth magnetic field or the misalignment between the magnetic axis of the coil and the rotation axis of the suspended bar during the oscillation of the bar. De Haas imagined a simple solution for this second problem: directly coil the iron bar with a conducting wire. The system, iron bar plus coil, oscillates together without possibility of misalignment between their axes (Fig3.b).

The artefact found at Ampère's Museum (Fig 3.c1) corresponds to the de Haas experiment. It is a coiled bar having the same length of 23 cm wounded with enameled copper wire 0.08 mm thick as described in the de Haas paper of 1915. In the upper part of the device (Fig3.c2), the suspension wire is fixed on top and two lateral branches drive thin wires to feed the coil with minimum friction not to affect the rotation. The bottom section of the device (Fig3.c3), incorporates a ballast potentially employed for mechanical stabilization and resonance frequency tuning. Following four months of intensive research, we discovered the artefact in 2023, only to find it absent from the Ampère Museum's inventory. We found letters attesting to the donation by Geertruida de Haas-Lorentz with the agreement of her husband in 1959. The donation, two photos of Einstein and W.J. de Haas (Fig 1) and the artefact, was registered in the 1962 published *Bulletin de la Société des Amis d'André-Marie Ampère*. The Boerhaave Museum in Leiden also holds an element also donated by Geertruida de Haas-Lorentz corresponding exactly to the central part of the artefact found at the Ampère's Museum, i.e., the iron core with its cooper coil and having exactly the same dimensions. De Haas may have fabricated spare pieces or multiple elements to test different set-ups. The donation of the artefact to the Ampere Museum may find a link with the fact that W.J. de Haas was member of the Patronage Committee of the *Société des Amis d'André-Marie Ampère*, society created in 1930 in order to run the Ampère Museum. The Ampère Museum (Fig1.d) was inaugurated in the following year and was labeled EPS Historic site in 2021.

As a final remark, during their experiments on "molecular currents", Ampère as well as Einstein and de Haas, can be seen as being on the verge of identifying groundbreaking new phenomena. Ampère nearly preempted Faraday's discovery of electromagnetic induction by almost a decade. Likewise, Einstein's and de Haas' work should be seen as a precursor to the detecting of the electron's spin the following decade, but further experiments and theoretical work had to be in place before a better understanding of the electron's spin could be established. These instances illustrate how the zeal to corroborate initial assumptions can lead to a too narrowed focus. They are therefore beautiful examples of the epistemological richness and complexity of the dialog between theory and experiments.

For readers interested on more details and history of the Einstein-de Haas experiments we refer to several published works[9,3,10]